# Colossal magnetocapacitive effect in differently synthesized and doped CdCr$_2$S$_4$

S. Krohns[a], F. Schrettle[a], P. Lunkenheimer[a,*], V. Tsurkan[a,b], A. Loidl[a]

[a]Experimental Physics V, Center for Electronic Correlations and Magnetism, University of Augsburg, D-86135 Augsburg, Germany
[b]Institute of Applied Physics, Academy of Sciences of Moldova, MD-2028 Chişinău, R.. Moldova

ABSTRACT

In the present work, we address the question of an impurity-related origin of the colossal magnetocapacitive effect in the spinel system CdCr$_2$S$_4$. We demonstrate that a strong variation in the dielectric constant below the magnetic transition temperature or in external magnetic fields also arises in crystals prepared without chlorine. This excludes that an inhomogeneous distribution of chlorine impurities at the surface or in the bulk material gives rise to the unusual effects in the spinel multiferroics. In addition, we show that colossal magnetocapacitive effects can be also generated in chlorine-free ceramic samples of CdCr$_2$S$_4$, doped with indium.

*Corresponding author.
E-mail address: peter.lunkenheimer@Physik.Uni-Augsburg.de

## 1. Introduction

The investigation of multiferroic materials, showing the simultaneous occurrence of ferroelectric and magnetic order, is a highly active field of research [1,2,3]. Especially the strong magnetocapacitive effects exhibited by some of these materials could pave the way to many appealing applications in modern electronics, e.g., for new advances in data-storage techniques. Recently, our group has reported the occurrence of a multiferroic state and colossal magnetocapacitance in various spinel systems, including CdCr$_2$S$_4$ (CCS) [4,5], CdCr$_2$Se$_4$ (CCSe) [6] and HgCr$_2$S$_4$ (HCS) [7].

This is not unexpected as spinels are prone to polar order and magnetocapacitive effects (see, e.g., [8,9,10,11]). However, while there is certainly unequivocal evidence for strong magnetocapacitive effects in these materials, the suggested polar order is a matter of debate and its microscopic origin is unclear. HCS exhibits spiral spin order at low temperatures [12] and probably can be treated along the same theoretical footing as other spiral magnets [13,14]. However, CCS and CCSe are ferromagnetic semiconductors and short range polar order evolves above the ordering temperature. In addition, no soft phonons have been found experimentally [15] or in LDA calculations [16]. However, strong charge-transfer processes have been detected in FIR spectroscopy [15] and indications for a local loss of inversion symmetry were reported from Raman experiments [17]. In addition, instead of soft phonons, electromagnons seem to be the relevant excitations in multiferroics [18]. In any case, it is clear that the polar order in the ferromagnetic spinels cannot be explained along conventional routes and unconventional scenarios may come into play as considered, e.g., in Refs. [10] and [19].

In most multiferroics, electrical polarization is hard to prove due to a relatively high conductivity and due to suspected local order. When assessing the experimental results, one has to be aware that strong contributions to the dielectric properties of materials also can arise from interfacial polarization effects. It is known since long [20] and was found, e.g., by our group in many different materials [21,22] that they can lead to so-called Maxwell-Wagner (MW) relaxations, which can give rise to very high apparent values of the dielectric constant and strong frequency dependence of the dielectric properties. Under certain circumstances, also magnetocapacitive behavior can arise from such effects [23,24] and consequently interfacial polarization was considered as an alternative explanation of the magnetocapacitance in CCS [5,23,25] and HCS [7,26] (for some arguments against such a view, see Refs. [27] and [28]). By measurements with different contact types, we could exclude external contacts leading to the observed phenomena [4,6]. Other possible origins of MW relaxations could be a non-stoichiometric surface layer [26] or a heterogeneous distribution of impurities [25], both arising from the chlorine [29] that was used as a transport agent during crystal preparation. Indeed, in [26] unpublished results from a different group on HCS crystals grown without chlorine were quoted having revealed no dielectric anomalies. Another problem is the fact that in first measurements of polycrystalline samples of CCS, again prepared without chlorine, no magnetocapacitive effect was found [30].

Thus, in the present work we have experimentally checked the proposed explanation in terms of chlorine impurities by performing further investigations on samples prepared via different routes. Especially, we provide results on single crystals of CCS prepared with bromine as transport agent, which, compared to chlorine, is much less likely to enter the lattice. In addition, polycrystals with indium admixture, $Cd[Cr_{1-x}In_x]_2S_4$ with $x = 0.05$, also prepared without any chlorine were investigated. In both cases, magnetocapacitive effects, very similar to those obtained for the crystals prepared by chlorine transport are observed. This finding demonstrates that a Maxwell-Wagner mechanism due to chlorine impurities [25,26] cannot explain the magnetocapacitive effect in the sulpho spinels.

## 2. Experimental details

Single crystals of CCS were grown by chemical transport reaction using bromine as the transport agent at 850°C resulting in samples in form of regular octahedrons with edge lengths up to 2 mm. Wave-length dispersive microanalysis of the samples revealed almost ideal stoichiometry within the accuracy of method (0.2 % for cations and 1% for sulfur). No bromine doping was detected. As-grown single crystals were annealed in vacuum at a temperature between 500 and 750°C during 0.5 - 2 hours. Polycrystalline samples with substitution of indium for chromium in the range 1-10 mol% were prepared by conventional solid state synthesis at 850°C. The sample investigated in the present work was annealed in sulfur atmosphere at 700 °C for 24 - 48 hours and in vacuum at 750 °C for 48 - 150 hours. Similar results were also obtained for ceramics without vacuum annealing. X-ray analysis showed single phase compositions without any impurity phases both in as-grown and annealed samples. Indium is known to enter the spinel structure mainly at the Cr site, although at high substitution level a small amount of In at the Cd site can be also found. For the dielectric measurements, a platelike sample was prepared from the as-grown crystal using conventional polishing technique with diamond paste applied to both sides. Opposite sides of the single crystalline sample (area $1.2 \times 1.5$ mm$^2$, thickness 0.5 mm) and the ceramic sample (diameter 6 mm, thickness 1.2 mm) were covered with silver paint. For details on the dielectric measurement techniques, see Refs. [4] and [31]. The typical soft-ferromagnetic properties, reported by us for samples prepared with chlorine transport [4], are well reproduced also by the new single crystals and an identical magnetic ordering temperature of $T_c = 84.5$ K is found.

## 2. Results and discussion

Figure 1(a) shows the temperature dependence of the dielectric constant $\varepsilon'$ of the chlorine-free CCS single-crystal for selected frequencies. Just as for the crystals grown with chlorine transport [4,5], at temperatures below $T_c$ a strong upturn of $\varepsilon'(T)$ shows up, pointing to a considerable coupling of the magnetic and dielectric properties. At $T > T_c$, $\varepsilon'(T)$ shows an increase, shifting to higher temperatures for higher frequencies, and a tendency to saturate at values of the order of 1000. This finding indicates relaxation-like behavior, which, however, is not as well pronounced as in the previously investigated samples. It seems that the additional increase in $\varepsilon'(T)$, observed at $T > 200$ K in Refs. [4,5] and ascribed to contact effects [4,6], is stronger in the present sample and superimposes the relaxation feature. Also the conductivity $\sigma'$ shown in Fig. 1(b) exhibits a strong anomaly at $T_c$, qualitatively similar to the behavior reported in [5]. However, quantitative deviations occur, especially at $T > T_c$ where the frequency dependence of $\sigma'$ is much weaker than in the Cl prepared crystals [5]. To understand the origin of these differences, measurements in a broader frequency range in further crystals and the analysis of the frequency dependence is necessary, which is out of the scope of the present work.

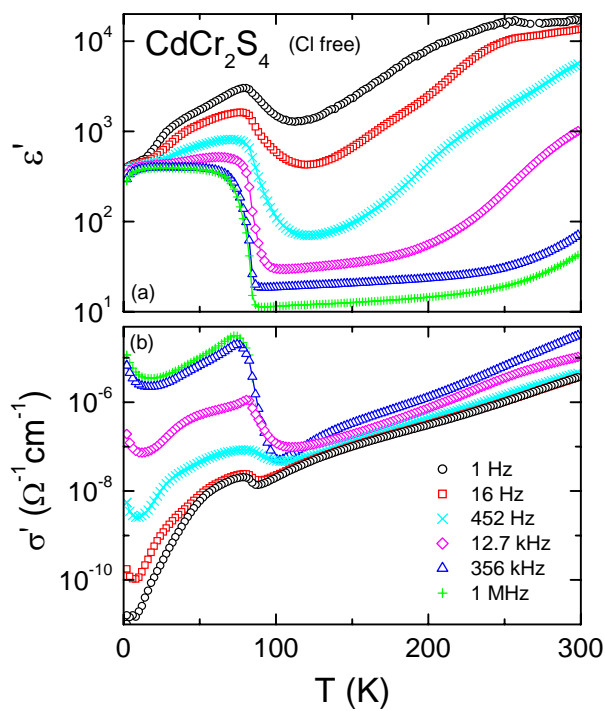

Fig. 1. Temperature dependence of the dielectric constant (a) and conductivity (b) at various frequencies measured at a CCS single crystal prepared without chlorine.

In Fig. 2, the temperature-dependent dielectric constant for two frequencies is shown, measured under an external magnetic field of 7 T and compared to the results with zero field. Clearly, there is a significant magnetocapacitive effect with a strong magnetic-field induced increase of $\varepsilon'$ at temperatures around $T_c$. It is of similar magnitude as reported by us for the crystals grown by chlorine transport [4,5].

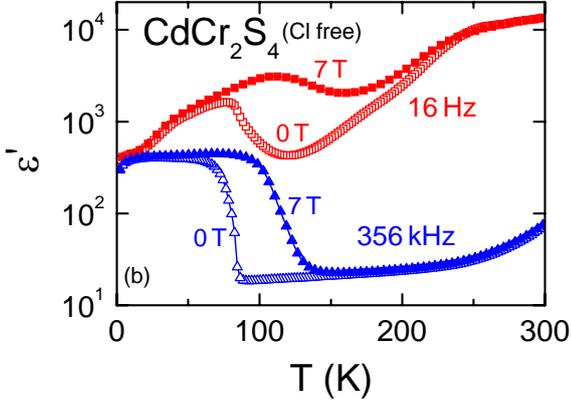

Fig. 2. Temperature dependence of the dielectric constant at two selected frequencies (open symbols), shown together with $\varepsilon'(T)$ measured in an external magnetic field of 7 T (closed symbols).

Thus, overall one can state that in these new samples, prepared using bromine as the transport agent, the same type and magnitude of magnetocapacitive effects are found as for crystals prepared with chlorine [4]. The ionic radius of chlorine (1.81 Å) [32], used as transport agent in our earlier investigation, rather closely matches that of sulfur (1.84 Å) [32] and therefore there is some likeliness of chlorine impurities. This is not the case for bromine, its ion radius (1.96 Å) [32] being significantly larger. Indeed, an investigation using wavelength dispersive electron-probe microanalysis revealed no indication of bromine in our samples. Thus an inhomogeneous impurity distribution of the transport agent as proposed in Refs. [25] and [26] cannot be the cause of the magnetocapacitive effects in CCS.

As mentioned in the introduction, so far the magnetocapacitive effect in CCS has not been observed in polycrystalline samples [30] Motivated by the fact that In-doped single crystals of CCS show strongly enhanced effects [30] we have prepared In-doped ceramic samples. In Fig. 3, the temperature dependence of the dielectric constant and the conductivity at various frequencies is shown for one of these samples, doped with 5% indium. Again, a clear rise of $\varepsilon'(T)$ is observed when the sample enters the magnetically ordered state. In Fig. 3(a), at high temperatures the signature of relaxational behavior shows up. A close inspection reveals two plateau values of the order of 10 and 1000. One may speculate that the relaxation leading to the upper plateau is of Maxwell-Wagner type and due to external contacts or grain boundaries, while the lower one is of intrinsic nature. To check for this notion, further experiments are necessary. The conductivity [Fig. 3(b)] is in rather good quantitative agreement with the one obtained in single crystals prepared with Cl [5]. Corresponding to the two plateaus in $\varepsilon'$, at $T > T_c$ two shoulders shifting with frequency are observed in $\sigma'(T)$, again pointing to two different relaxations.

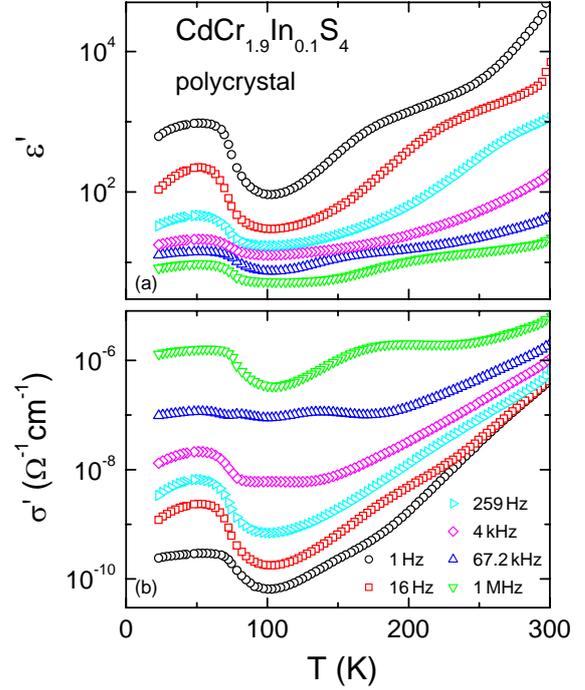

Fig. 3. Temperature dependence of the dielectric constant (a) and conductivity (b) at various frequencies for In-doped CCS ceramic sample.

It should be noted that a magnetocapacitive effect in the newly prepared single crystals was obtained after annealing the samples in vacuum as described above. This is quite in contrast to the first set of crystals [4,5,6,30], where annealing in vacuum suppressed the magnetocapacitive effect [30,33]. In the polycrystals the effect shows up for In-doped samples at doping levels in the range of 3 - 7 %. It is observed with and without vacuum annealing after the first sintering procedure in sulfur atmosphere. Subsequent sulfur annealing suppresses the effect.

## 4. Conclusion

In summary, obviously the magnetocapacitive effect in CCS is highly sensitive to details of sample preparation and doping and the exact role of stoichiometry and defects still has to be resolved. Of course, we do not claim that the present results finally prove the intrinsic nature of the magnetocapacitive behavior of CCS. However, they clearly exclude one possible non-intrinsic mechanism, namely the heterogeneous distribution of impurities as considered in [25] and [26], and thus one has to think of other intrinsic or non-intrinsic explanations. The provided recipe for preparing also ceramic samples showing the effect, hopefully will trigger more experimental work investigating the intriguing magnetocapacitive properties of the thio-spinels.

**Acknowledgements**

This work was supported by the Deutsche Forschungsgemeinschaft via the Sonderforschungsbereich 484.